\documentclass[epj]{svjour}
\usepackage{amssymb,graphics,graphicx,amsmath,epsfig}

\begin{document}

\authorrunning{P. Fleischer et al}
\title{The two-proton shell gap in Sn isotopes}

\author{
 P. Fleischer\inst{1},
 P. Kl\"upfel\inst{1},
 T. Cornelius\inst{2}, 
 T.J. B\"urvenich\inst{3}, 
 S. Schramm\inst{2}
 J.A. Maruhn\inst{2}, 
 P.--G. Reinhard\inst{1},
}
\institute{
         Institut f{\"u}r Theoretische Physik,
        Universit{\"a}t Erlangen,
        Staudtstrasse 7, D-91058 Erlangen, Germany
    \and
  Institut f\"ur Theoretische Physik, Universit\"at Frankfurt,
  D-60325 Frankfurt, Germany
    \and
Theoretical Division, Los Alamos National Laboratory, Los Alamos,
New Mexico 87545
}
\abstract{
We present an analysis of two-proton shell gaps in Sn isotopes. 
As the theoretical tool we use self-consistent mean-field models, namely
the relativistic mean-field model and the Skyrme-Hartree-Fock approach,
both with two different pairing forces, a delta interaction (DI) model
and a density-dependent delta interaction (DDDI).
We investigate the influence of nuclear deformation as well as 
collective correlations and find that both effects contribute
significantly. 
Moreover, we find a further significant dependence on the
pairing force used. The inclusion of 
deformation plus correlation effects and the use of
DDDI pairing provides agreement with the data.
}

\maketitle

\section{Introduction}
Understanding nuclear shell structure has been a key issue of nuclear
physics for decades \cite{Goe49a,Hax49a}. It remains a topic of large
current interest in connection with nuclei far from the valley of
stability for which a large pool of new data is now available
%\cite{gsi-masses}. The predictive value of nuclear structure models in
\cite{Rad00a}. The predictive value of nuclear structure models in
the various regimes of exotic nuclei depends very much on their
ability to describe shell structure in quantitative detail,
particularly for superheavy nuclei with their subtle dependence on
shells \cite{Rei02ar,Ben01a}. On the other hand, there is the basic
problem that we do not dispose of observables which give direct
experimental access to the single particle levels of nuclei. There is
the seemingly ``direct'' deduction through separation energies and
excitation spectra of neighboring odd nuclei. The proper modeling of
these observables by mean-field models, however, is rather involved as
it requires inclusion of a proper blocking description, polarization
effects and breaking of time-reversal symmetry \cite{Ber80a,Rut98}. It
is desirable to have complementing information on the single particle
structure and one hopes that data deduced only from even-even nuclei
give simpler access to spectral gaps. In this context the two-nucleon
shell gaps are often considered\cite{Ben99a}.  They are simply the
second differences of binding energies and thus easily available from
experiment. The workload comes on the theoretical side because this
quantity, being a difference of large numbers, is extremely sensitive
to all sorts of corrections and thus requires careful modeling.
Systematic measurements on the long chain of Pb isotopes
\cite{Rad00a,Nov02a} have shown a steady decrease of the two-proton
shell gap towards the proton drip line indicating some shell
quenching. It was found in a subsequent theoretical analysis that
deformation masks the data such that the two-proton shell gap shows a
quenching while the spectral gap does not \cite{Ben02}. Proper
inclusion of deformed mean fields brought satisfactory agreement with
the experimental results.  It is the aim of the present paper to
continue these investigations for a different test case, namely the
two-proton shell gap in the chain of Sn isotopes. It will turn out
that deformation effects are as important as they were in the case of
the Pb isotopes, but we will find also that ground-state deformation
alone is insufficient to explain the experimental two-proton shell
gaps in Sn.  Thus we have extended the studies to include a new
aspect, namely the effect of collective ground-state correlations on
the two-proton shell gaps. We will consider two brands of mean-field
models, namely the relativistic mean field model (RMF) as well as the
non-relativistic Skyrme-Hartree-Fock (SHF) approach and a variety of
parameterizations within both models.  For technical reasons, the
study of ground-state correlations is confined to SHF models.

The paper is outlined as follows:
In section \ref{sec:frame}, we very briefly explain the formal
framework, the mean-field models and the treatment of collective
correlations.
In section \ref{sec:res}, we present and discuss the results,
where subsection \ref{sec:deformed} is concerned with deformed
mean-field states and subsection \ref{sec:gsc} with the impact
of ground-state correlations.

\section{Framework}
\label{sec:frame}
\subsection{Self-consistent mean-field models}
Our investigation is performed in the framework of self-consistent
mean-field theories, namely the non-relativistic Skyrme-Hartree-Fock
approach (SHF) and the relativistic mean-field (RMF) model, for the
formal details and an extensive discussion of their properties see
\cite{Ben03aR}. Moreover, we employ for the RMF two slightly different
brands, namely the traditional variant with finite-range meson
exchange (RMF-FR) \cite{Ser86aR,Rei89} and the more recent version
employing point couplings (RMF-PC) \cite{nhm,buer02}. This span of
models explores different physical ingredients: a comparison of SHF
with RMF tests the non-relativistic versus relativistic approach, the
comparison of RMF-FR and RMF-PC tests the importance of finite-range
mean fields.
In each of these models, there exists a great variety of
different parameterizations which all deliver a comparable and
excellent description of bulk properties in stable nuclei but can
differ substantially in the realm of exotic nuclei or when looking at
more subtle observables, as we will do here. One thus has to use in
such investigations a representative sample of different
parameterizations to disentangle genuine mean-field effects from 
particularities of a given parameterization.
From the SHF family we will consider: SkM$^*$ as a widely used
traditional standard \cite{Bar82a}, Sly6 as a recent fit which
includes information on isotopic trends and neutron matter
\cite{Cha97a}, SkI3 as a fit which maps the relativistic iso-vector
structure of the spin-orbit force and takes care of the surface
thickness \cite{Rei95a}, and SkO \cite{Rei99b} as a recent fit in
similar fashion as SkI3 but with bias on a larger effective mass and a
better adjusted asymmetry energy.
For the RMF model, we consider the RMF-FR parameterizations
NL-Z2\cite{Ben00a} and NL3\cite{Gam90a} as well as the RMF-PC force
PC-F1 \cite{buer02}. Both NL-Z2 and PC-F1 have been fitted to a
similar set of data as SkI3 and SKO, including information on the
nuclear charge formfactor. NL3 has been adjusted with particular
emphasis on isovector properties and incompressibility.
For all models, pairing is added at the level of BCS with
Lipkin-Nogami correction. 
We use a zero-range delta interaction (DI) 
$V_{\rm pair}=V^{\rm(DI)}_\nu\delta(r_1-r_2)$ 
as pairing force, and as alternative recipe
the density-dependent delta-interaction (DDDI) \cite{Ton79a,Kri90a,Ter95a}
$V_{\rm pair}=
 V^{\rm(DDDI)}_\nu\delta(r_1-r_2)\left[1-\rho(\bar{r})/\rho_{0}\right]$.
In both cases, $\nu$ stands for protons or neutrons.  The pairing
strengths $V^{\rm(DI)}_\nu$ or $V^{\rm(DDDI)}_\nu$ are adjusted such
that the average pairing gaps 
$\bar\Delta=\sum\alpha u_\alpha v_\alpha \Delta_\alpha/
 \sum\alpha u_\alpha v_\alpha$
\cite{Ben00b} are fitted to the pairing gaps from the experimental
odd-even staggering of binding energies in a few representative
semi-magic nuclei. Actually, we use Sn isotopes for the neutron gaps,
namely
$\Delta_n(^{112}{\rm Sn})=1.41\,{\rm MeV}$,
$\Delta_n(^{120}{\rm Sn})=1.39\,{\rm MeV}$,
$\Delta_n(^{124}{\rm Sn})=1.31\,{\rm MeV}$,
and some $N=82$ isotones for the proton gaps 
$\Delta_n(^{136}{\rm Xe})=0.98\,{\rm MeV}$, $\Delta_n(^{144}{\rm
Sa})=1.25\,{\rm MeV}$.  The adjustment is done for each force
separately because the much different effective masses call for
different pairing strengths in each case. The gaps are well fitted in
the average (better than 1\%). The trends are reproduced within at
least 10\% precision.
\begin{table}
\begin{center}
\begin{tabular}{lcccccc} \hline\noalign{\smallskip}
force                & 
%$\rho_{\rm nm}$ fm$^{-3}$   &
%$E/A$   [MeV]     &
%$K_\infty$   [MeV]        &
%$m_0^*/m$            &
%$a_{\rm sym}$ [MeV] &
%$a'_{\rm sym}$ [MeV] &
$\rho_{\rm nm}$    &
$\frac{E}{A}$      &
$K_\infty$      &
$\frac{m^*}{m}$            &
$a_{\rm sym}$  &
$\kappa_{\rm TRK}^{\mbox{}}$ 
%\\
% &{\footnotesize {}[fm$^{-3}$]} &{\footnotesize [MeV]} &{\footnotesize  [MeV]} & 
%& {\footnotesize [MeV]} & {\footnotesize[MeV]} & 
\\ \noalign{\smallskip}\hline\noalign{\smallskip}
   SkM* &  0.160&  -15.8&   217&  0.79&   30&  0.53\\
   SLy6 &  0.159&  -15.9&   230&  0.69&   32&  0.25\\
   SkI3 &  0.158&  -16.0&   258&  0.58&   34&  0.25\\
   SkO4 &  0.161&  -15.8&   224&  0.90&   32&  0.17\\
\noalign{\smallskip} \hline
NL-Z2 & 0.151&  -16.2&   173&  0.58&    42&  0.72\\
NL3   & 0.148&  -16.2&   272&  0.60&    37&  0.68\\
PC-F1 & 0.151&  -16.2&   270&  0.61&    38&  0.70
\\ \noalign{\smallskip}\hline\noalign{\smallskip}
\end{tabular}
\end{center}
\caption{\label{tab:inm}
Nuclear matter properties for the considered forces:
saturation density $\rho_{\rm nm}$ in units of fm$^{-3}$, 
binding energy $E/A$ in units of MeV,
incompressibility $K_\infty$ in units of MeV, 
isoscalar effective mass $m^*/m$, 
symmetry energy $a_{\rm sym}$ in units of MeV,
and 
sum-rule enhancement factor $\kappa_{\rm TRK}$ (equivalent
to isovector effective mass).
}
\end{table}
The basic properties of these forces in terms of nuclear matter
%\marginpar{\#1}
parameters are given in table \ref{tab:inm}.  A detailed discussion is
found in \cite{Ben03aR}.  There are the typical systematical differences
between RMF and SHF for the symmetry energy and for the sum rule
enhancement $\kappa_{\rm TRK}^{\mbox{}}$. These may be related to
different slopes in the spherical shell gaps as seen in forthcoming
figures.  Note, in particular, the large span of effective masses for
SHF which cover the range from 0.58 to 0.9 and which is the most
important variation because the effective mass has an influence on the
spectral gap as well as on deformation and correlation properties.

\subsection{Collective correlations}
\label{sec:GSCcomp}

The Cd and Te isotopes far away from the magic neutron shell $N\!=\!82$ turn
out to be extremely soft in quadrupole deformations. Their ground state goes
beyond a pure mean-field description. It is, in fact, a coherent superposition
of mean-field states at various deformations. This means that we include some
correlations, namely those associated with low-energy collective motion. The
effective energy functionals can still be used for that task as low-energy
collective motion can be derived as the adiabatic limit of time-dependent mean
fields \cite{Bar78a,Goe78a,Rei87aR}. As practical procedure, we employ the
generator-coordinate method (GCM) with Gaussian overlap approximation (GOA),
see e.g. \cite{Rei87aR,Bon90b}. In fact, we use a variant of GOA which takes
care of the topology of coupled rotations and quadrupole vibrations
\cite{Rei78b,Hag02,Hag03}. We summarize here briefly the basic ingredients.
Details are found in \cite{FleischerDiss,Fle04a}.

A series of collectively deformed mean-field states $|\Phi_q\rangle$
is generated by quadrupole constrained Skyrme-Hartree-Fock where $q$
stands for the actual quadrupole momentum.  Their energy expectation
value ${\cal V}(q)=\langle\Phi_q|\hat{H}|\Phi_q\rangle$ provides a raw
collective potential in quadrupole space.  The path does also define
the collective momentum as generator for deformation
$\hat{P}_q|\Phi_q\rangle=\imath\partial_q|\Phi_q\rangle$.  Collective
masses and moments of inertia are computed by linear response to
deformation, i.e.  
%\marginpar{\#2}
%
${\cal M}_q^{-1}=\langle[\hat{R}_q,[\hat{H},\hat{R}_q]]\rangle$ 
where $\hat{R}_q$ is defined from linear response to $\hat{P}_q$, i.e. $[\hat{H},\hat{R}_q]\propto\hat{P}_q$, and similarly to rotation
(ATDHF cranking). Zero-point energies $E^{\rm(ZPE)}$
for vibrations and rotations are
computed from the fluctuations in quadrupole
$\langle\hat{P}_q^2\rangle$ and angular momentum
$\langle\hat{J}\rangle$ together with the associated masses. These
constitute quantum corrections to the collective potential
\cite{Rei75a,Rei76a,Rei89}. The true collective potential $V$
is then obtained from the raw potential ${\cal V}$ as
$$
  V
  =
  {\cal V}
  -
  E^{\rm(ZPE)}
  \quad.
$$
All ingredients together finally yield a
generalized collective Bohr-Hamiltonian
\begin{eqnarray*}
    \hat{H}^{\rm(coll)} 
    &=&
    -\frac{1}{\beta^4}\partial_\beta B\beta^4\partial_\beta 
    - \frac{1}{\beta^2\,\sin3\gamma}\,\partial_\gamma 
      B_\gamma\,\sin3\gamma\,\partial_\gamma 
\\
  &&
    + \sum_{k=1}^3 \frac{\hat{L}_k^{\prime 2}}{2\Theta_k}
    \qquad +V
\end{eqnarray*}
where the potential $V$ and masses $B,\Theta$ are functions of
deformation $\beta,\gamma$.  The dependence on triaxiality $\gamma$ is
obtained by interpolation of axial results, a procedure which is
justified for the nearly spherical nuclei considered here.
The ground-state solution in
collective space then represents the collectively correlated
ground-state energy. The physical collective zero-point energy is, of
course, positive, but the quantum corrections which are subtracted
from the raw potential are larger, so that at the end we obtain a more
tightly bound correlated ground state.
%

%
%
%
%%%%%%%%%%%%%%%%%%%%%%%%%%%%%%%%%%%%%%%%%%%%%%%%%%%%%%%%%%%
\section{Results and discussion}\label{sec:res}
%%%%%%%%%%%%%%%%%%%%%%%%%%%%%%%%%%%%%%%%%%%%%%%%%%%%%%%%%%%
%

\subsection{The effect of ground-state deformation}
\label{sec:deformed}

The mean-field description of nuclei provides the full details of the
single-nucleon energies $\epsilon_k$ as the eigenvalues of the single-particle
Hamiltonian. With this information at hand, shell effects can be easily
quantified in various manners, e.g. as the spectral gap which is the energy
difference between the highest occupied state and the lowest unoccupied state
(called HOMO-LUMO gap in molecular physics) or as the shell correction energy
\cite{strut67,brack72} (for a recent analysis in super-heavy elements see
e.g. \cite{Ben03aR}). There is, however, no direct experimental access to
single nucleon spectra due to rearrangement and core polarization effects
\cite{Ber80a,Rut98}.
There remains as a fairly simple and experimental criterion the
two-nucleon shell gaps from which we discuss here in particular the
two-proton shell gaps
%
%\begin{subequations}
%\begin{eqnarray*}
$$
  \delta_{2p}(Z,N) 
  = 
  \left(E(Z\!-\!2) - 2E(Z) +E(Z\!+\!2)\right)\Big|_N
$$
%\end{eqnarray*}
%\end{subequations}
%
where $|_N$ means that all energies are taken at the same neutron
number $N$.  The $\delta_{2p}$ are just the second differences of
binding energies. That observable is close to twice the spectral gap
in the nucleus with $(N,Z)$ provided that the mean field does not
undergo substantial changes from one nucleus to the next.  In an
earlier publication, we had investigated the two-proton shell gap in
the chain of Pb isotopes and we found that deformation softness can
indeed lead to strong changes in the mean field which, in turn, modify
the two-nucleon gaps \cite{Ben02}. This helped to clarify a puzzle:
the experimental two-proton gaps hint a ``shell quenching'' towards
the proton drip line while spherical 
%\marginpar{\#4}
mean-field calculations show
always a large and robust spectral gap. The quenching seen for the
$\delta_{2p}$ is an effect of deformation popping up in the step from
$Z\!=\!50$ to the neighbors with $Z\!=\!50\!\pm\!2$. Proper inclusion
of ground-state deformation delivered nice agreement between
mean-field calculations and data. We are now going to explore such
effects for the chain of Sn isotopes.
\begin{figure*}
%\centerline{\epsfig{figure=gle_figures/figure2.eps,width=16.4cm}}
\centerline{\epsfig{figure=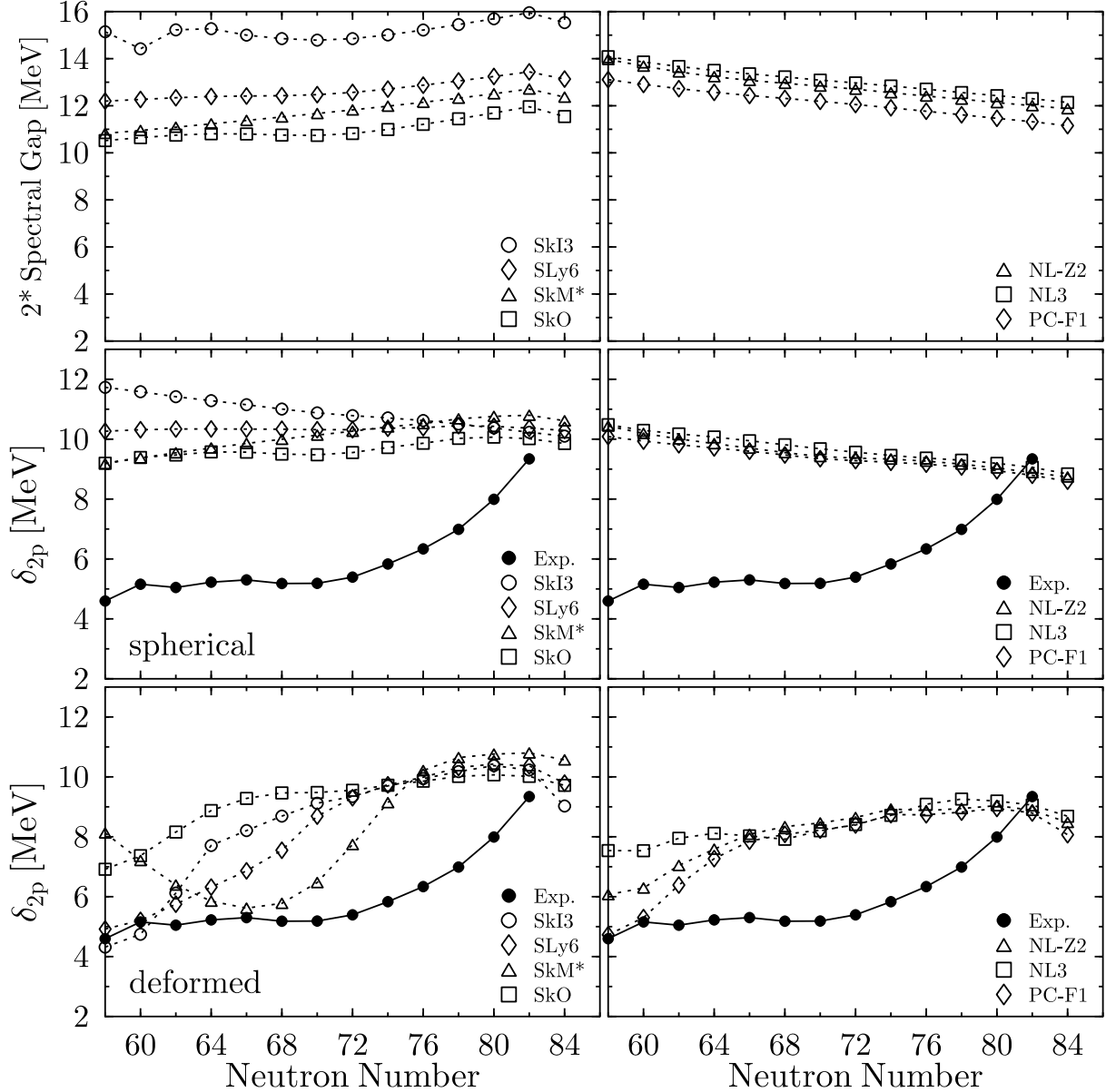,width=16.4cm}}
\caption{\label{tin_sph_def}
Upper panels: twice the spectral gap for protons (difference between
lowest unoccupied and highest occupied proton state) along the chain
of Sn isotopes.  Middle panels: Two-proton shell gaps $\delta_{2p}$
calculated in the spherical mean field.  Lower panels: Two-proton shell
gaps $\delta_{2p}$ calculated in the axially deformed mean field.  The
left panels shows results from various SHF parameterizations as
indicated and the right panels from RMF.
}
\end{figure*}

Figure \ref{tin_sph_def} summarizes results from spherical and
deformed mean-field calculations.  The uppermost panels show the
spectral gaps in Sn isotopes. Within the SHF forces they show a clear
dependence on the effective mass: the gaps increase with decreasing
$m^*/m$. The comparison with RMF hints that the symmetry energy has an
impact on the trend with neutron number. However, the gaps from RMF
are a bit lower than expected from the very low effective masses in
RMF. The reasons for that are still unclear.
%\marginpar{\#3}

The middle panels of figure \ref{tin_sph_def} show the $\delta_{2p}$
from spherical mean-field calculations. At first glance, they behave
much similar to the spectral gap (upper panels), as expected.  There
are, however, changes in quantitative detail due to spherical
polarization and rearrangement effects. The values are generally down
shifted by 1-2 MeV and the dependence on $m^*/m$ is reduced.
Nonetheless, these corrections still stay sufficiently small
\cite{Rut98} to leave $\delta_{2p}$ as approximate measure of the
spectral shell gaps. On the other hand the spherical $\delta_{2p}$ are
far from the data and the deviation increases with decreasing neutron
number.
%
% The deviation varies amongst the different mean-field models. All RMF
% forces provide about the same results in sizes and trends while Skyrme
% forces show more variations. The force SkI3 shares the trend of the
% RMF results but has a generally larger spherical $\delta_{2p}$. The
% other Skyrme forces come closer to RMF in value but differ visibly in
% their trend.  Note that the Skyrme forces have an isovector spin-orbit
% which differs from the RMF, except for SkI3. Moreover, all Skyrme
% forces here have significantly lower asymmetry energy than the forces
% from the RMF family.  Both features probably contribute to the
% differences seen for the spherical results.
%

\begin{figure*}
%\centerline{\epsfig{figure=gle_figures/figure3.eps,width=16.4cm}}
\centerline{\epsfig{figure=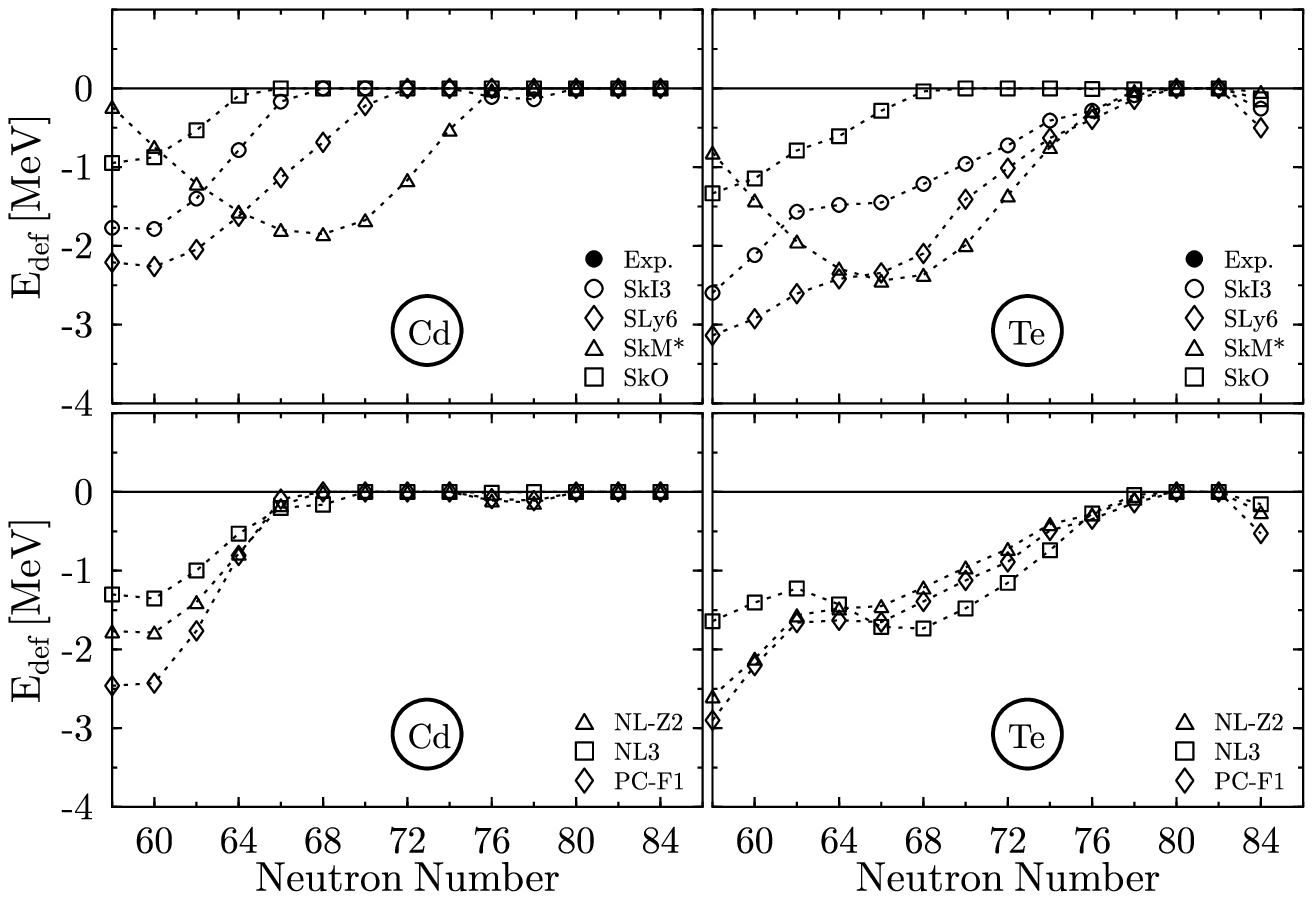,width=16.4cm}}
\caption{\label{deformation}
The deformation energy $E^{\rm(d)}$ 
in the neighboring isotopic chains Cd and Te for the various forces as
indicated.}
\end{figure*}

The lowest panels of figure \ref{tin_sph_def} show results from
axially deformed mean-field calculations.  The deformation when active
lowers the $\delta_{2p}$ because the neighboring isotones gain binding
energy through deformation while Sn remains spherical. In order to
illustrate the mechanism, we separate the energies into (leading)
spherical part and the contribution from deformation:
\begin{eqnarray*}
  \delta_{2p}(Z,N) 
  &=&
  \delta_{2p}^{(0)}(Z,N)
  + 
  \delta_{2p}^{\rm(d)}(Z,N) 
  \quad,
\\
  \delta_{2p}^{(0)}(Z,N)
  &=& 
  \left(E^{(0)}(Z\!-\!2) - 2E^{(0)}(Z) +E^{(0)}(Z\!+\!2)\right)\Big|_N
  \quad,
\\
  \delta_{2p}^{\rm(d)}(Z,N)
  &=& 
  \left(E^{\rm(d)}(Z\!-\!2)  +E^{\rm(d)}(Z\!+\!2)\right)\Big|_N
  <0
  \quad.
\end{eqnarray*}
$E^{(0)}$ is the energy of the spherical configuration. 
$E^{\rm(d)}$ is the gain in binding through deformation, thus
$E^{\rm(d)}<0$. The Sn isotopes, the intermediate chain, stay spherical while
the softer neighbors Cd and Te develop deformation. 
This produces the unique direction
for the deformation effect on $\delta_{2p}$ for Sn.
The detailed deformation energies for the considered nuclei are shown
in figure \ref{deformation}.  Comparing that with the lowest panels of
figure \ref{tin_sph_def} confirms that the reduction of the
$\delta_{2p}$ around $N \approx 60$ originates from the deformation of
the neighboring nuclei.
All forces show the deformation effect towards the side of lower
$N$, but the onset of this effect differs, mainly due to the
differences in effective mass \cite{Blu89a}. The lower $m^*/m$ the
later (in terms of low $N$) the onset. The RMF results, for example,
deform only for low $N$ and stay closer to each other what the onset
of deformations is concerned because they have all about the same low
$m^*/m$. The case of SkO is an exception from that rule. Note that
this force was fitted with somewhat different conditions,namely a
strong constraint on the two-neutron shell gap in $^{208}$Pb
\cite{Rei99b}. This seems to have deep consequences in the shell
structure with side-effects as observed here, and in forthcoming
figures.  Related to the shifted onsets is a shift of the minima in
the curves.  The first force showing up with deformations, SkM$^*$, is
also the first to return to sphericity when approaching the lower
neutron shell closure at $N=50$. Thus we see a clear maximum of
deformation effects within the plotting window. The other forces will
bend up as well, but outside the plotting window.
In spite of the variations in the results, all models still
overestimate the two-proton shell-gap in the region $N=72..80$ as seen
clearly in the lower panels of figure \ref{tin_sph_def}.

\subsection{The impact of ground-state correlations}
\label{sec:gsc}

A possible reason for that may be ground-state correlations (GSC). Nuclei
near the onset of deformation are usually soft against quadrupole 
fluctuations even
if the mean deformation still stays close to sphericity. These quadrupole quantum
fluctuations in the ground state lead to a correlation which also produces
extra binding. The Sn nuclei are more rigid (owing to the proton shell 
closure $Z=50$) than
their neighbors and thus we can expect a further lowering of the $\delta_{2p}$
particularly in the transitional region.
\begin{figure}
\unitlength1mm
%\begin{picture}(50,0)
%\centerline{\includegraphics[width=8.5cm,angle=-0]{Sn_chain_d2p_ski3.eps}}
\centerline{\includegraphics[width=8.5cm,angle=-0]{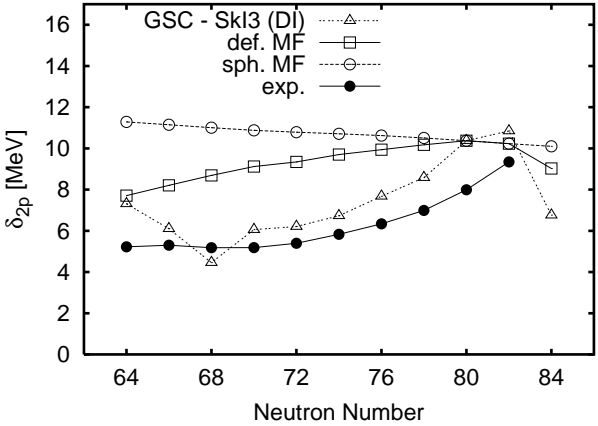}}
%\end{picture}
\caption[zu]{\label{pic:ski3-d2p}\small 
  The two-proton shell gap $\delta 2{\rm p}$ 
  calculated for the interaction SkI3 with spherical mean field,
  deformed mean field, and with collective GSC
  from quadrupole fluctuations (computed as sketched
  in section \ref{sec:GSCcomp}). All cases used DI pairing.
  The experimental data is taken from \protect \cite{Rad00a}}
\end{figure}
The effect on the $\delta_{2p}$ is demonstrated in figure
\ref{pic:ski3-d2p} for the force SkI3 as test case.  We again see the
significant step towards the experimental trend when switching from
spherical to deformed mean-field calculations. The next step is to account
for the GSC and this indeed provides the expected additional lowering
of the $\delta_{2p}$.  The effect is particularly pronounced in the
transitional region where the deformation effect alone is still too
small. However, it is not yet large enough to reach agreement with the
data. As an aside, we note the somewhat strange detail at $N=82$ where
the correlation effect even changes sign. The nucleus $^{132}$Sn is
doubly magic and we suspect that the simple pairing treatment is a bit
risky in the immediate vicinity of a double shell closure, and it is
just this vicinity which contributes to the $\delta_{2p}$. On the other
hand, the effect remains rather small and well localized. It stays
away from the transitional region which is the focus of our
discussions.

\begin{figure}
\unitlength1mm
\centerline{\includegraphics[width=8.5cm,angle=-0]{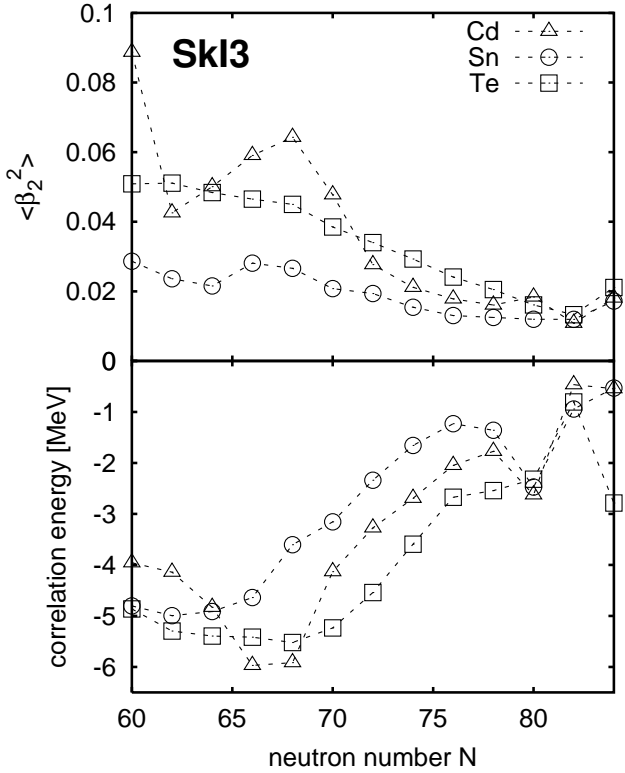}}
\caption[zu]{\label{pic:corE}
  Details for the correlation effect with SkI3 and DI pairing.
  Upper panel: The squared quadrupole expectation value
  $\langle\hat{\beta}_2^2\rangle$ for Sn and its even neighbors.
  Lower panel: The correlation energy (difference between
  fully correlated ground-state energy and the deformed
  mean-field minimum).
}
\end{figure}
%
%\marginpar{\#6}
Complementing information on the correlation effect is provided in
figure \ref{pic:corE}. The upper panel shows the squared quadrupole
expectation value $\langle\hat{\beta}_2^2\rangle$ for the correlated
state. It embraces the ground-state deformation as well as the
collective fluctuations. The fluctuation part is to a good
approximation proportional to the B(E2) values. The total
$\langle\hat{\beta}_2^2\rangle$ is directly related to the correlation
effect on the r.m.s. radii \cite{Rei79a,Bar85b}. There is, however,
not such a direct relation to the correlation energy as can be seen
from comparison with the lower part of the figure. The reason is that
the correlation energy is composed from two counteracting
contributions, the negative quantum correction energy  (see section
\ref{sec:GSCcomp}) and the positive collective zero-point energy.
The $\langle\hat{\beta}_2^2\rangle$ is small for the semi-magic Sn
isotopes while it grows systematically towards mid-shell for the
neighboring elements Cd and Te. This shows that collective vibrations
are much softer for these nuclei.  
The correlation energy, on the other hand, grows towards mid-shell for
all three elements. But it is larger  for Cd and Te in the
transitional region such that finally the correlations just serve to
fill the gap between deformed results (lower panel in figure 
\ref{tin_sph_def}) and experiment.

\begin{figure}
\unitlength1mm
%\centerline{\includegraphics[width=8.5cm,angle=-0]{Sn_chain_d2p_forces.eps}}
\centerline{\includegraphics[width=8.5cm,angle=-0]{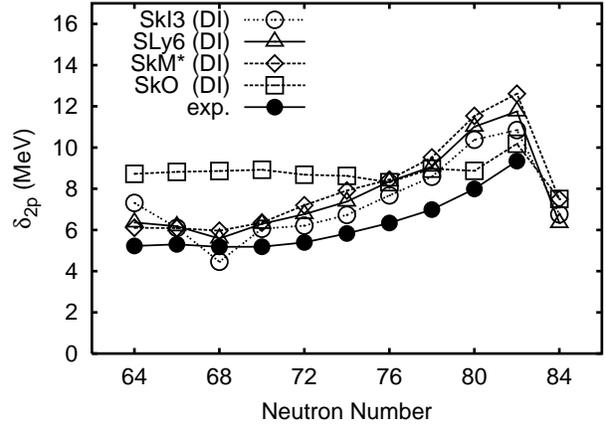}}
\caption[zu]{\label{pic:all-d2p}
  Computed $\delta 2p$ including the quadrupole Ground state
  Correlations (GSC) for the three Skyrme interactions SkI3, SLy6 and
  SkM* all using delta-interaction (DI) pairing.  The experimental
  data are taken from \protect \cite{Rad00a}}
\end{figure}
In order to exclude that this may be a particular problem of the force
SkI3 (used for figure \ref{pic:ski3-d2p}) in figure
\ref{pic:all-d2p} results with GSC and for a broader variety of Skyrme
forces are shown. The force SkO (having a rather large effective mass
$m^*/m=0.9$) is even farther away from the data, even when including
deformation and GSC as done here. The other forces yield on the average
the same result as we had seen for SkI3 before.  It is interesting to
note that the actual size of the correlation effect differs amongst
the forces as can be seen from the fact that the ordering of the
results in relation to data is different as in the deformed mean field
shown in figure \ref{tin_sph_def}. The force SkI3 is a bit special as
it shows some fluctuations at neutron number $N=68$ which are appear
only when correlations are included. This is due to a small sub-shell
closure which exists for SkI3 and not for the other three forces.  In
spite of the observed variances of the results, none of them reaches
fully near the data in a systematic manner.

\begin{figure}
\unitlength1mm
%\centerline{\includegraphics[width=8.5cm,angle=-0]{Sn_chain_d2p_forces.eps}}
\centerline{\includegraphics[width=8.5cm,angle=-0]{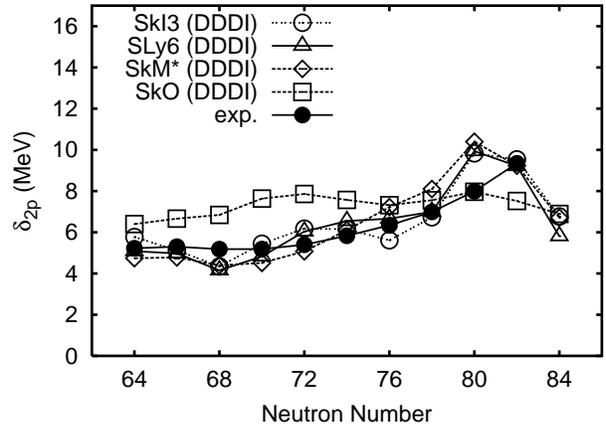}}
\caption[zu]{\label{pic:all-d2p-DDDI}
  As figure \ref{pic:all-d2p}, but with DDDI pairing.
}
\end{figure}
Thus far we have considered a certain variation of Skyrme forces.
Pairing is the other crucial ingredient in the effective energy
functional. Up to now, we have considered only the DI pairing.  A
widely used alternative is the surface, or DDDI, pairing.  Figure
\ref{pic:all-d2p-DDDI} shows results of calculations with ground-state
correlations using the same Skyrme forces as above, but now with DDDI
pairing.  The force SkO behaves again a bit strange. The alternative
pairing recipe helps a bit but cannot counterweight the obviously
inappropriate mean-field background of this particular
parameterization.  The main effect is that DDDI pairing reduces the
two-proton shell gap by about one more MeV and this brings the results
for the three well performing forces (SkM$^*$, SLy6, SkI3) on top of
the experimental data.  The effect of DDDI versus DI pairing is only
about 1 MeV for that subtle observable $\delta_{2p}$, but crucial for
final success.  The difference comes mainly from the neighboring
isotones Cd and Te which are a bit less soft with DDDI. The example is
one hint that DDDI pairing may be more realistic, but much more
evidence has yet to be collected to support such a statement, see
e.g. the many discussion of that question from different aspects in
%\marginpar{\#5}
\cite{Avd99,Dob02a,Dug04a}.

\section{Conclusions}
We have investigated the two-proton shell gap in Sn isotopes in the
framework of self-consistent mean-field models, the
Skyrme-Hartree-Fock approach as well as the relativistic mean-field
model. In a first step beyond pure mean field, we have also considered
the effect of collective correlations stemming from quadrupole
vibrations.  In order to work out systematic trends, we have
considered a variety of different parameterizations in both mean-field
models.

In a first step, we have investigated the effect of ground-state
deformations on the two-proton shell gap. In both trend
and order of magnitude the results were quite similar to those of a 
previous study on Pb
isotopes thus corroborating that effect as a general feature for all
nuclei. The deformation effects are negligible only near the doubly
magic nuclei and take over as soon as the neutron number is
sufficiently far from a magic shell closure. 

Different from our earlier studies on Pb isotopes, the resulting
two-proton shell gaps did not match the experimental data even when
including the deformation effects. Thus we have also checked the
impact of collective ground-state correlations. They do indeed lower
the two-proton shell gap in the critical region by about 1-2 Mev and
thus bring the results closer to the data.  It was then found that a
further crucial ingredient is the pairing model.  We compared a delta
interaction (DI) pairing with a density-dependent delta interaction
(DDDI). There is again about 1 MeV difference in the results. The DDDI
pairing together with the collective ground-state correlations
finally delivers a perfect agreement with the experimental data for a
variety of different Skyrme forces.

In summary, the results show that the two-proton shell gap is a very
subtle observable which is extremely sensitive to various details of
the treatment, spherical and deformed polarization effects as well as
correlations. The final result reflects different features of the
forces as, e.g., the effective mass (which defines the underlying
spectral gap) and the various response parameters (which enter the
polarization effects). The two-nucleon shell gaps certainly provide
useful information about nuclear shell structure. But the situation is
similarly involved as with single-nucleon information from odd nuclei.
It requires a careful consideration of all ingredients to compare
theory with experimental data.

\bigskip

\noindent
Acknowledgment:
The authors thank M. Bender for many inspiring discussions and
helpful remarks. 
This work was supported
in part by the Bundesministerium f\"ur Bildung und Forschung (BMBF),
Project Nos.\ 06 ER 808 and 06 ER 124.

% bib 
\bibliographystyle{prsty}
\bibliography{delta_2p}

\begin{thebibliography}{10}

\bibitem{Goe49a}
M. G{\"o}ppert-Mayer, Phys. Rev. {\bf 75},  1969  (1949).

\bibitem{Hax49a}
O. Haxel, J.~H.~D. Jensen, and H.~E. Suess, Phys. Rev. {\bf 75},  1766  (1949).

\bibitem{Rad00a}
T. Radon, H. Geissel, G. M{\"u}nzenberg, B. Franzke, T. Kerscher, F. Nolden,
  Y.~N. Novikov, Z. Patyk, C. Scheidenberger, F. Attallah, K. Beckert, T. Beha,
  F. Bosch, H. Eickhoff, M. Falch, Y. Fujita, M. Hausmann, F. Herfurth, H.
  Irnich, H.~C. Jung, O. Klepper, C. Kozhuharov, Y.~A. Litvinov, K.~E.~G.
  L{\"o}bner, F. Nickel, H. Reich, W. Schwab, B. Schlitt, M. Steck, K.
  S{\"u}mmerer, T. Winkler, and H. Wollnik, Nucl. Phys. A {\bf 677},  75
  (2000).

%\bibitem{gsi-masses}
%T. Radon {\em et al}, Nucl. Phys. A {\bf 677},  75  (2000).

\bibitem{Rei02ar}
P.-G. Reinhard, M. Bender, and J.~A. Maruhn, Comm. Mod. Phys. A {\bf 2},  177
  (2002).

\bibitem{Ben01a}
M. Bender, W. Nazarewicz, and P.-G. Reinhard, Phys. Lett. B {\bf 515},  42
  (2001).

\bibitem{Ber80a}
V. Bernard and {Nguyen Van Giai}, Nucl. Phys. A {\bf 348},  75  (1980).

\bibitem{Rut98}
K. Rutz, M. Bender, P.-G. Reinhard, J.~A. Maruhn, and W. Greiner, Nucl. Phys.
  {\bf A634},  67  (1998).

\bibitem{Ben99a}
M. Bender, K. Rutz, P.-G. Reinhard, J.~A. Maruhn, and W. Greiner, Phys. Rev. C
  {\bf 60},  034304  (1999).

\bibitem{Nov02a}
Y.~N. Novikov, F. Attallah, F. Bosch, M. Falch, H. Geissel, M. Hausmann, T.
  Kerscher, O. Klepper, H.-J. Kluge, C. Kozhuharov, Y.~A. Litvinov, K.~E.~G.
  L{\"o}bner, G. M{\"u}nzenberg, Z. Patyk, T. Radon, C. Scheidenberger, A.~H.
  Wapstra, and H. Wollnik, Nucl. Phys. A {\bf 697},  92  (2002).

\bibitem{Ben02}
M. Bender, T. Cornelius, G. Lalazissis, J. Maruhn, W. Nazarewicz, and P.-G.
  Reinhard, Eur. Phys. J. {\bf A 14},  23  (2002).

\bibitem{Ben03aR}
M. Bender, P.-H. Heenen, and P.-G. Reinhard, Rev. Mod. Phys. {\bf 75},  121
  (2003).

\bibitem{Ser86aR}
B.~D. Serot and J.~D. Walecka, Adv. Nucl. Phys. {\bf 16},  1  (1986).

\bibitem{Rei89}
P.-G. Reinhard, Rep. Prog. Phys. {\bf 52},  439  (1989).

\bibitem{nhm}
B.~A. Nikolaus, T. Hoch, and D.~G. Madland, Phys. Rev. C {\bf 46},  1757
  (1992).

\bibitem{buer02}
T. B{\"u}rvenich, D.~G. Madland, J.~A. Maruhn, and P.-G. Reinhard, Phys. Rev. C
  {\bf 65},  044308  (2002).

\bibitem{Bar82a}
J. Bartel, P. Quentin, M. Brack, C. Guet, and H.-B. H{\aa}kansson, Nucl. Phys.
  {\bf A386},  79  (1982).

\bibitem{Cha97a}
E. Chabanat, P. Bonche, P. Haensel, J. Meyer, and R. Schaeffer, Nucl. Phys. A
  {\bf 627},  710  (1997).

\bibitem{Rei95a}
P.-G. Reinhard and H. Flocard, Nucl. Phys. {\bf A584},  467  (1995).

\bibitem{Rei99b}
P.-G. Reinhard, D.~J. Dean, W. Nazarewicz, J. Dobaczewski, J.~A. Maruhn, and
  M.~R. Strayer, Phys. Rev. C {\bf 60},  014316  (1999).

\bibitem{Ben00a}
M. Bender, Phys. Rev. C {\bf 61},  031302(R)  (2000).

\bibitem{Gam90a}
Y.~K. Gambhir, P. Ring, and A. Thimet, Ann. Phys. N. Y. {\bf 198},  132
  (1990).

\bibitem{Ton79a}
F. Tondeur, Nucl. Phys. {\bf A315},  353  (1979).

\bibitem{Kri90a}
S.~J. Krieger, P. Bonche, H. Flocard, P. Quentin, and M.~S. Weiss, Nucl. Phys.
  {\bf A517},  275  (1990).

\bibitem{Ter95a}
J. Terasaki, P.-H. Heenen, P. Bonche, J. Dobaczewski, and H. Flocard, Nucl.
  Phys. {\bf A593},  1  (1995).

\bibitem{Ben00b}
M. Bender, K. Rutz, P.-G. Reinhard, and J. Maruhn, Eur. Phys. J. A {\bf 8},  59
   (2000).

\bibitem{Bar78a}
M. Baranger and M. V{\'e}n{\'e}roni, Ann. Phys. N. Y. {\bf 114},  123  (1978).

\bibitem{Goe78a}
K. Goeke and P.-G. Reinhard, Ann. Phys. N. Y. {\bf 112},  328  (1978).

\bibitem{Rei87aR}
P.-G. Reinhard and K. Goeke, Rep. Prog. Phys. {\bf 50},  1  (1987).

\bibitem{Bon90b}
P. Bonche, J. Dobaczewski, H. Flocard, P.-H. Heenen, and J. Meyer, Nucl. Phys.
  {\bf A510},  466  (1990).

\bibitem{Rei78b}
P.-G. Reinhard, Z. Physik {\bf A285},  93  (1978).

\bibitem{Hag02}
K. Hagino, P.-G. Reinhard, and G. Bertsch, Phys. Rev. C {\bf 65},  064320
  (2002).

\bibitem{Hag03}
K. Hagino, G. Bertsch, and P.-G. Reinhard, Phys. Rev. C {\bf 68},  024306
  (2003).

\bibitem{FleischerDiss}
P. Fleischer, Ph.D. thesis, Universit{\"a}t, Erlangen / Germany, 2003.

\bibitem{Fle04a}
P. Fleischer, P. Kl\"upfel, P.-G. Reinhard, and J.~A. Maruhn, preprint
  (unpublished).

\bibitem{Rei75a}
P.-G. Reinhard, Nucl. Phys. {\bf A252},  120  (1975).

\bibitem{Rei76a}
P.-G. Reinhard, Nucl. Phys. {\bf A261},  291  (1976).

\bibitem{strut67}
V. Strutinski, Nucl.Phys. {\bf A95},  420  (1967).

\bibitem{brack72}
M. Brack, J. Damg{\aa}rd, A. Jensen, H. Pauli, V. Strutinski, and C. Wong, Rev.
  Mod. Phys. {\bf {\bf 44}},  320  (1972).

\bibitem{Blu89a}
V. Blum, J. Fink, P.-G. Reinhard, J.~A. Maruhn, and W. Greiner, Phys. Lett.
  {\bf 223B},  123  (1989).

\bibitem{Rei79a}
P.-G. Reinhard and D. Drechsel, Z. Phys. {\bf A290},  85  (1979).

\bibitem{Bar85b}
F. Barranco and R.~A. Broglia, Phys. Lett. {\bf B151},  90  (1985).

\bibitem{Avd99}
A. Avdeenkov and S. Kamerdzhiev, JETP Lett. {\bf 69},  715  (1999).

\bibitem{Dob02a}
J. Dobaczewski, W. Nazarewicz, and M. Stoitsov, Eur. Phys. J {\bf 15},  21
  (2002).

\bibitem{Dug04a}
T. Duguet, Phys. Rev. C {\bf 69},  054317  (2004).

\end{thebibliography}

\end{document}